\documentclass[preprint,nofootinbib,amsmath,amssymb,aps,preprintnumbers,superscriptaddress]{revtex4-1}
\usepackage{graphicx} % Required for inserting images
\usepackage{amsmath,amsthm,latexsym,amscd,amsfonts,bm,hyperref,empheq}
\usepackage{here}
\usepackage{mathrsfs}
\usepackage{xcolor}

\begin{document}

\preprint{KEK-TH-2866, KEK-Cosmo-0433, KOBE-COSMO-26-06}

\title{Graviton Floor}
%\author{teruaki.suyama.sheep }

\author{Himeka Matsuo}
%\email[]{himeka.matsuo@grad.nao.ac.jp}
\affiliation{Astronomical Science Program, Graduate Institute for Advanced Studies, SOKENDAI, 2-21-1 Osawa, Mitaka, Tokyo 181-8588, Japan}
\affiliation{National Astronomical Observatory of Japan, 2-21-1 Osawa, Mitaka, Tokyo 181-8588, Japan}

\author{Asuka Ito}
%\email[]{asuka@phys.sci.kobe-u.ac.jp}
\affiliation{Department of Physics, Kobe University, Kobe 657-8501, Japan}

\author{Kazunori Kohri}
%\email[]{kazunori.kohri@gmail.com}
\affiliation{National Astronomical Observatory of Japan, 2-21-1 Osawa, Mitaka, Tokyo 181-8588, Japan}
\affiliation{Department of Astronomy, The University of Tokyo, Bunkyo-ku, Hongo, Tokyo 113-0033, Japan}
\affiliation{Astronomical Science Program, Graduate Institute for Advanced Studies, SOKENDAI, 2-21-1 Osawa, Mitaka, Tokyo 181-8588, Japan}
\affiliation{Theory Center, IPNS, KEK, 1-1 Oho, Tsukuba, Ibaraki 305-0801, Japan}
\affiliation{Kavli IPMU (WPI), UTIAS, The University of Tokyo, Kashiwa, Chiba 277-8583, Japan}

\author{Teruaki Suyama}
%\email[]{suyama@phys.sci.isct.ac.jp}
\affiliation{Department of Physics, Institute of Science Tokyo, 2-12-1 Ookayama, Meguro-ku, Tokyo 152-8551, Japan}

\author{Ryutaro Tomomatsu}
%\email[]{tomomatsu.r.b02d@m.isct.ac.jp}
\affiliation{Department of Physics, Institute of Science Tokyo, 2-12-1 Ookayama, Meguro-ku, Tokyo 152-8551, Japan}

%\date{October 2025}

\begin{abstract}
It has been observed that the Universe is permeated by the cosmic photon background,
ranging from radio waves to gamma rays.
We investigate the conversion of the photon background into gravitons in the presence of background magnetic fields in the Milky Way Galaxy and in blazar jets.
We find that the resulting graviton background is dominated by the contribution generated in blazar jets.
Importantly, this graviton background constitutes a graviton floor for high-frequency gravitational wave detectors searching for new physics, analogous to the neutrino floor.
\end{abstract}

\maketitle

\section{Introduction}

Magnetic fields are ubiquitous in the Universe, from small scales like stars to galactic and even cosmological scales~\cite{Widrow:2002ud}.
They influence astrophysical processes such as star formation, cosmic ray acceleration, and the dynamics of galaxies and galaxy clusters.
The Gertsenshtein effect is a phenomenon in which gravitons and photons convert into each other in the presence of external magnetic fields~\cite{Gertsenshtein:1962,Raffelt:1987im}.
Recently, this graviton–photon conversion has been studied in the context of high-frequency gravitational wave searches~\cite{Pshirkov:2009sf,Dolgov:2012be,Domcke:2020yzq,Ramazanov:2023nxz,Liu:2023mll,Ito:2023fcr,Ito:2023nkq,Dandoy:2024oqg,He:2023xoh,Lella:2024dus,McDonald:2024nxj,Kushwaha:2025mia,Li:2025eoo,Amaral:2026bef,Baker:2026dqf}, where the converted photons can serve as a probe of high-frequency gravitational waves.
Possibilities for probing high-frequency gravitational waves using graviton–photon conversion in geomagnetic fields~\cite{Liu:2023mll,Ito:2023nkq,Baker:2026dqf}, pulsar magnetospheres~\cite{Ito:2023fcr,Dandoy:2024oqg,McDonald:2024nxj}, galactic magnetic fields~\cite{Ito:2023nkq,Lella:2024dus}, intergalactic magnetic fields~\cite{Domcke:2020yzq,Ito:2023nkq,He:2023xoh,Kushwaha:2025mia,Li:2025eoo}, and blazar jets~\cite{Matsuo:2025blj} have been studied.

In this paper, we consider the inverse process, namely the conversion of photons into gravitons.
It has been observed that the Universe is permeated by the cosmic photon background, ranging from radio waves to gamma rays~\cite{Hill:2018trh}.
These photons necessarily convert into gravitons in the presence of background magnetic fields, so that a graviton background is formed, typically in the frequency range above $\sim 10$ MHz.
The possibility of generating a graviton background due to the presence of primordial magnetic fields has been investigated in the literature~\cite{Chen:1994ch,Cillis:1996qy,Chen:2013gva,Fujita:2020rdx}.
In this work, we first investigate the conversion of photons into gravitons within the Milky Way Galaxy, where the properties of magnetic fields have been well observed~\cite{Haverkorn:2014jka,Boulanger:2018zrk,Haverkorn:2004fw,Iacobelli:2013fqa,Haverkorn:2008tb}.
We also study the conversion process in blazar jets.
Blazar jets are known to host magnetic fields whose strengths can be inferred from observations of their spectral energy distributions (SEDs)~\cite{Fan_2023}.
By accumulating converted gravitons in each blazar jet, we evaluate the produced graviton background.
Importantly, the resultant graviton background constitutes a graviton floor for high-frequency gravitational wave detectors searching for new 
physics~\cite{Aggarwal:2025noe,Ito:2019wcb,Ito:2020wxi,Berlin:2021txa,Ito:2022rxn,Ito:2023bnu,Berlin:2023grv,Bringmann:2023gba,Kanno:2023whr,Ito:2025mgm,Cai:2025fpe}, 
analogous to the neutrino floor~\cite{Billard:2013qya,OHare:2021utq}.

The organization of the paper is as follows.
In Sec.~\ref{Grav-photon}, we review graviton–photon conversion in background magnetic fields.
In Sec.~\ref{Milky}, we study the conversion of the cosmic photon background into gravitons in the Milky Way Galaxy.
In Sec.~\ref{sec_blazar}, we study the conversion of photons in blazar jets using the observed photon spectra, and evaluate the resulting graviton background.
The final section is devoted to the conclusion.

\section{Graviton-photon mixing in magnetic fields}\label{Grav-photon}
In this section, we give a short review of the graviton–photon conversion
under a background magnetic field~\cite{Gertsenshtein:1962,Raffelt:1987im}.
We consider the following action:
\begin{align}\label{action}
  S=\int d^4x\sqrt{-g} \left[ \frac{M^2_{{\rm pl}}}{2}R-\frac{1}{4}F_{\mu\nu}F^{\mu\nu} \right],
\end{align}
where $M_{{\rm pl}}$ denotes the reduced Planck mass, $R$ is the Ricci scalar, and $g$ is the determinant of the metric $g_{\mu\nu}$.
The field strength of the electromagnetic field is defined as
$F_{\mu\nu} = \partial_\mu \mathscr{A}_\nu - \partial_\nu \mathscr{A}_\mu$,
where $\mathscr{A}_\mu$ is the vector potential.

Let us write the vector potential and the metric in the forms
\begin{align}
  \mathscr{A}_{\mu}(x)&=\bar{A}_{\mu}+A_{\mu}(x),\\
  g_{\mu\nu}(x)&=\eta_{\mu\nu}+\frac{2}{M_{{\rm pl}}}h_{\mu\nu}(x),
\end{align}
where $\bar{A}_{\mu}$ corresponds to the background magnetic field through
$\bar{B}^i = \epsilon^{ijk}\partial_j \bar{A}_k$,
$\eta_{\mu\nu}$ denotes the Minkowski metric, 
and $h_{\mu\nu}(x)$ is a traceless-transverse tensor.
In the following, we choose the gauge $A_0 = 0$ and keep only the two transverse photon modes,
since the longitudinal mode (if any) is not relevant for our purpose.
We assume that gravitons and photons propagate along the $z$-direction.
The background magnetic field is taken along the $y$-direction, orthogonal to the propagation,
namely $\boldsymbol{\bar{B}} = (0, \bar{B}, 0)$.
Then, the polarization bases for the vector and tensor fields can be chosen as
\begin{align}
\label{bases}
  e^+_i=
  \begin{pmatrix}
  1 \\
  0 \\
  0
  \end{pmatrix},\enspace
  e^\times_i=
  \begin{pmatrix}
  0 \\
  1 \\
  0
  \end{pmatrix},\enspace
  \epsilon^+_{ij}=\frac{1}{\sqrt{2}}
  \begin{pmatrix}
  1 & 0 & 0 \\
  0 & -1 & 0 \\
  0 & 0 & 0
  \end{pmatrix},\enspace
  \epsilon^\times_{ij}=\frac{1}{\sqrt{2}}
  \begin{pmatrix}
  0 & 1 & 0 \\
  1 & 0 & 0 \\
  0 & 0 & 0
  \end{pmatrix}.
\end{align}
Here, two linear polarizations of photons are labeled by $+$ and $\times$ according to 
plus and cross modes of gravitational waves.
The electromagnetic field and the gravitational wave can be expanded as follows
\begin{align}\label{A}
  A_i&= \sum_{\sigma} e^{-i(\omega t-kz)}A^\sigma(z)e^\sigma_i \, , \\\label{h} 
  h_{ij}&= \sum_{\sigma} e^{-i(\omega t-kz)}h^\sigma(z)\epsilon^\sigma_{ij} \, ,
\end{align}
with the bases (\ref{bases}) for $\sigma = +, \times$.
The quantities $A^\sigma(z)$ and $h^\sigma(z)$ denote the amplitudes of the electromagnetic wave and the gravitational wave, respectively.
These amplitudes can change during propagation because of the graviton–photon mixing.
By using Eqs.~(\ref{action})–(\ref{h}) and including the effects of magnetized plasmas as well as higher-order QED corrections, one obtains~\cite{Ito:2023fcr}

\begin{align}\label{Ahplusmode}
  \left[i\partial_z+
  \begin{pmatrix}
  M_+ & -i\frac{\bar{B}(z)}{\sqrt{2}M_{{\rm pl}}}
\\
  i\frac{\bar{B}(z)}{\sqrt{2}M_{{\rm pl}}} & 0 
  \end{pmatrix}
  \right]
  \begin{pmatrix}
    A^+(z)\\
    h^+(z)
  \end{pmatrix}
  \simeq0 , 
\end{align}
\begin{align}
  \label{Ahcrossmode}
  \left[i\partial_z+
  \begin{pmatrix}
  M_\times & -i\frac{\bar{B}(z)}{\sqrt{2}M_{{\rm pl}}} \\
  i\frac{\bar{B}(z)}{\sqrt{2}M_{{\rm pl}}} & 0 
  \end{pmatrix}
  \right]
  \begin{pmatrix}
    A^\times(z)\\
    h^\times(z)
  \end{pmatrix}
  \simeq0 ,
\end{align}
with
\begin{eqnarray}
 M_+ &=& -\sum_i \frac{1}{2\omega}\frac{\omega^2\omega_{p,(i)}^2}{\omega^2-\omega^2_{c,(i)}}+\frac{1}{2\omega}\frac{16\alpha^2\bar{B}(z)^2\omega^2}{45m_e^4}+\frac{1}{2\omega}\frac{88\pi^2\alpha^2T^4\omega^2}{2025m^4_e} , \\
 M_\times &=& -\sum_i \frac{\omega^2_{p,(i)}}{2\omega}+\frac{1}{2\omega}\frac{28\alpha^2\bar{B}(z)^2\omega^2}{45m_e^4}+\frac{1}{2\omega}\frac{88\pi^2\alpha^2T^4\omega^2}{2025m^4_e} .
\end{eqnarray}
In deriving Eqs.~(\ref{Ahplusmode}) and (\ref{Ahcrossmode}), we assumed that the scale of graviton-photon conversion is much larger than $k^{-1}$, and that photons are ultrarelativistic, i.e., $\omega \simeq k$. We also neglected the spatial derivatives of the magnetic field $\bar{B}$.
$T$ represents the temperature of the cosmic microwave background.
The plasma frequency $\omega_{p}$ and the cyclotron frequency $\omega_{c}$ are defined as
\begin{align}\label{pl}
  \omega_{p,(i)}&=\sqrt{\frac{4\pi\alpha n_{i}}{m_{i}\gamma_{i}}},\\
  \omega_{c,(i)}&=\frac{eB(z)}{m_i\gamma_{i}},
\end{align}
where $\alpha$ is the fine structure constant, $n_e$ ($n_p$) is the electron (proton) number density, and
$m_e$ ($m_p$) is the electron (proton) mass.  
$\gamma_e$ ($\gamma_p$) represents the Lorentz factor of electrons (protons), which
reduces the plasma frequencies because the effective masses of the plasma particles increase due to the Lorentz factors.
Although the mixing between electromagnetic waves and gravitational waves arises from the off-diagonal terms in Eqs.~(\ref{Ahplusmode}) and (\ref{Ahcrossmode}),
the diagonal terms, which give the effective photon masses, also affect the conversion efficiency.
Thus, not only the strength and configuration of the background magnetic field but also plasma parameters such as number densities need to be examined carefully.
When the magnetic field is not so strong,
the cyclotron frequency term in Eq.~(\ref{Ahplusmode}) can be neglected,
because the condition $\omega_c/\omega \ll 1$ is satisfied.
Under this assumption, the plus and cross modes experience nearly identical plasma effects.
When the cyclotron frequency is negligible, and in the low-frequency regime where the plasma contribution dominates over the higher-order QED effect,
the conversion rates of the plus and cross modes are essentially identical.
In contrast, at higher frequencies where the QED correction becomes more important,
the conversion efficiencies of the two modes can differ slightly.

\section{Gravitational wave production in the Milky Way Galaxy}\label{Milky}
In this section, we investigate the graviton-photon conversion within magnetic fields in the Milky Way Galaxy
and evaluate the produced gravitational waves from the cosmic photon background.
The typical magnetic field strength $B_{\rm G}$ lies in the range of $1\,\mu{\rm G}$--$10\,\mu{\rm G}$~\cite{Haverkorn:2014jka,Boulanger:2018zrk}.
There are two main components of the magnetic field: one is the large-scale magnetic field, which has
a large coherence length and a directional dependence following the Galactic spiral,
and the other is the small-scale magnetic field, whose coherence length is of order
$1\,\mathrm{pc}$--$100\,\mathrm{pc}$~\cite{Haverkorn:2004fw,Iacobelli:2013fqa,Haverkorn:2008tb}.
In this work, we focus on the latter component, which is assumed to have an isotropic random distribution.
Although an anisotropic random magnetic field component may also exist~\cite{Haverkorn:2014jka,Boulanger:2018zrk}, 
it is neglected here.
We note that graviton to photon conversion in small-scale magnetic fields was studied in~\cite{Ito:2023nkq}, while that in large-scale magnetic fields was studied in~\cite{Lella:2024dus}.
The conversion rates in these two components are comparable: which contribution dominates depends on the frequency, the magnitude and coherence length of the small-scale magnetic field, and the propagation direction relative to the large-scale magnetic field.
Therefore, in estimating the gravitational wave background produced in the Milky Way Galaxy
in a conservative manner, 
it would be sufficient to consider only the small-scale magnetic field component while varying 
the magnetic field strength.

Following~\cite{Ito:2023nkq},
we adopt a smoothly connected cellular model for the small-scale magnetic field distribution, in which
the magnetic field is taken to be homogeneous within domains of size $l_{\rm G}$, and
many such cells are assumed to fill the Milky Way Galaxy on scales of $\sim 10\,\mathrm{kpc}$.
The conversion probability between gravitons and photons in a single cell can be obtained by diagonalizing 
Eqs.~(\ref{Ahplusmode}) and (\ref{Ahcrossmode}).
After propagating through $N_{\rm G} \sim 10\,\mathrm{kpc}/l_{\rm G}$ cells, the total conversion probability
can be approximated as being multiplied by $N_{\rm G}$~\cite{Ito:2023nkq}, that is
\begin{equation}
  P^{(\sigma)} = N_{{\rm G}}
      \frac{l_{\rm os}^2 \tilde{B}_{{\rm G}}^{2}}
      {2M_{{\rm pl}}^{2}}
      \times
      \sin^{2}\left( \frac{l_{{\rm G}}}{l_{{\rm os}}} \right) , \label{pro}
\end{equation}
where we defined the oscillation length 
\begin{equation}
  l_{{\rm os}} = \frac{4\omega}{\sqrt{\left( \omega_{p,\sigma}^{2}  - \omega_{{\rm QED},\sigma}^{2}  
                 -  \omega_{{\rm CMB}}^{2} \right)^{2}
                 + \frac{8 \tilde{B}_{{\rm G}}^{2} \omega^{2}}{M_{{\rm pl}}^{2}} }}  .
\end{equation}
Note that we replace $B_{\rm G}$ with its averaged transverse component,
$\tilde{B}_{\rm G} = \sqrt{2/3}\,B_{\rm G}$.
When the oscillation length $l_{\rm os}$ is smaller than the magnetic-field coherence length $l_{\rm G}$,
the sine function in Eq.~(\ref{pro}) can be approximated by its average value, $1/2$.
Therefore, the conversion probability can be evaluated as follows:
\begin{widetext}
\begin{empheq}[left={P^{(\sigma)} \simeq \empheqlbrace}]{alignat=2}
  \frac{l_{\rm os}^2 \tilde{B}_{{\rm G}}^{2}}
      {4M_{{\rm pl}}^{2}}
 & \quad \quad {\rm for} \quad l_{{\rm os}} < l_{{\rm G}} \label{G1} , \\
\  N_{{\rm G}} \frac{l_{{\rm G}}^{2} \tilde{B}_{{\rm G}}^{2}}{2 M_{{\rm pl}}^{2}} 
       & \quad \quad {\rm for} \quad l_{{\rm os}} > l_{{\rm G}}  . \label{G2}
\end{empheq}
\end{widetext}
It is worth mentioning that, in Eq.\,(\ref{G2}),
we implicitly assumed that the direction of the magnetic field suddenly changes at the boundary of the cells, i.e., the variation of the magnetic field is non-adiabatic. 
If $l_{\rm os} \gg l_{\rm G}$, this assumption is justified.
In realistic situations, however, the transition between neighboring cells may be smooth, and the magnetic field can be regarded as an adiabatically varying background when 
$l_{\rm os} \ll l_{\rm G}$~\cite{Kartavtsev:2016doq,Ito:2023fcr,Ito:2023nkq}.
In such a case, there is no enhancement by a factor of $N_{\rm G}$.
Therefore, we replace $N_{\rm G}$ by unity in Eq.~(\ref{G1}).
%The oscillation length is shown in Fig.~\ref{osc}.
It should be noted that this treatment is conservative compared to studies adopting the hard cell model,
in which the factor $N_{\rm G}$ remains even for $l_{\rm os} < l_{\rm G}$.
%~\cite{Dolgov:2012be}.
%
%
\begin{figure}[th]
\centering
\includegraphics[width=12cm]{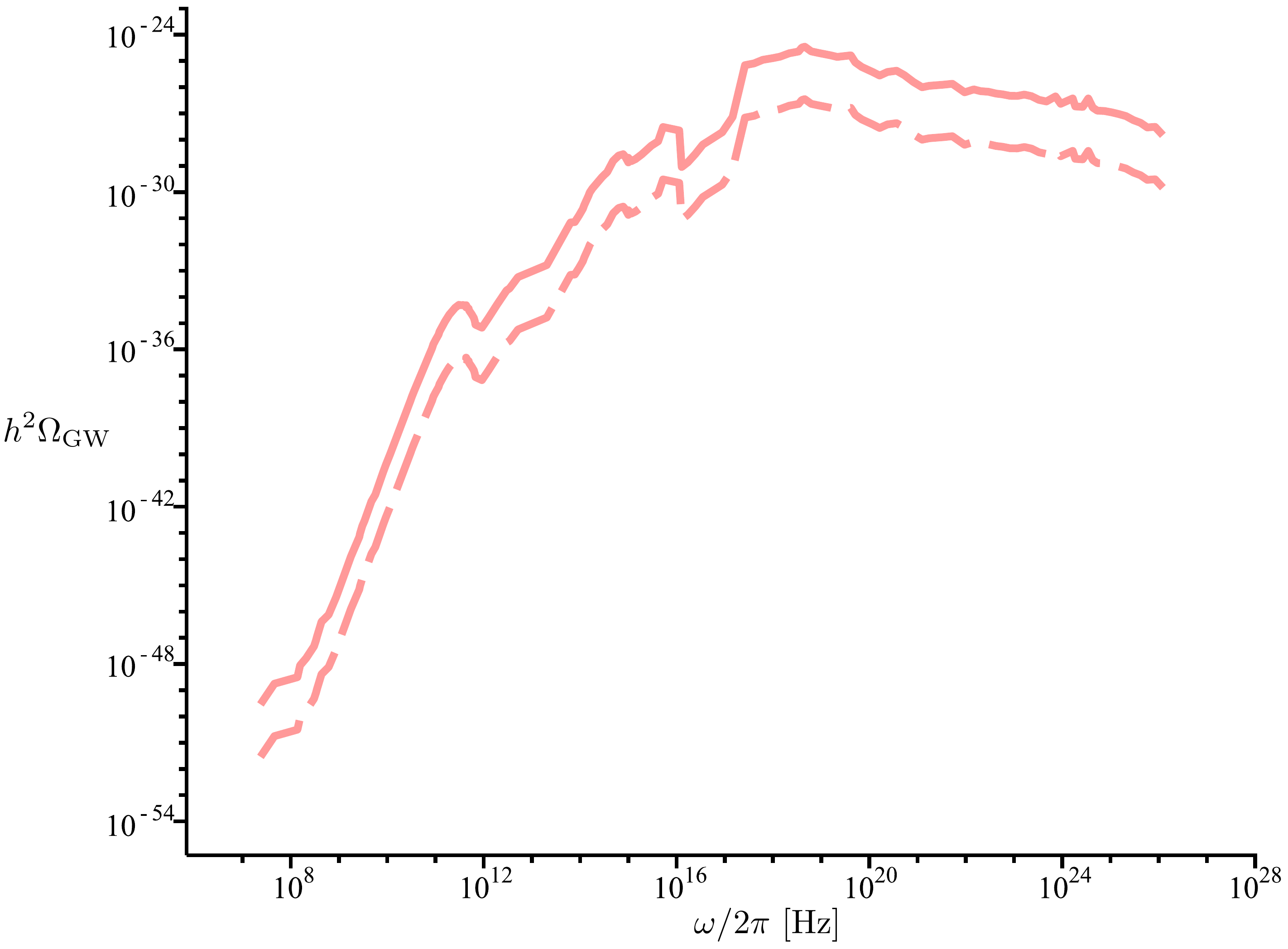}
\caption{Stochastic gravitational waves produced in the Milky Way Galaxy from the cosmic photon background.
The solid and dashed red lines correspond to magnetic field magnitudes of $10\mu$G and $1\mu$G,
respectively.
The coherence length of the magnetic field, i.e., the size of a cell, is assumed to be $100$pc.} 
\label{galaxy}
\end{figure}

Using the above method for calculating the conversion probability,
we evaluate the gravitational waves produced through 
photon–graviton conversion of background photons in the Milky Way Galaxy.
We approximate that the converted gravitational waves reach us isotropically.
The electron density varies with direction and distance from the Galactic center~\cite{Cordes:2002wz}.
Since a higher electron density generally decreases the conversion probability,
we adopt an electron density of $n_e = 7\times10^{-2}\,\mathrm{cm}^{-3}$~\cite{Cordes:2002wz}
to obtain a conservative estimate of the graviton-photon conversion probability.
%value of the electron density,
%We use the observed photon fluxes of
%the cosmic photon background~\cite{Hill:2018trh}
%to calculate the resulting stochastic gravitational wave background.
We use the observed photon fluxes of the cosmic photon background
compiled in~\cite{Hill:2018trh}, specifically the data shown in
Figs.~1--7 therein.
These photon fluxes are combined with the graviton-photon conversion
probability to calculate the resulting stochastic gravitational wave background.
The obtained gravitational wave abundance is shown in Fig.\,\ref{galaxy}.
In the evaluation, we assumed that the coherence length of the magnetic field, i.e., the size 
of a cell, is $100$pc. 
We note that, while the coherence length of the small-scale magnetic field ranges from $1$pc to $100$pc~\cite{Haverkorn:2004fw,Iacobelli:2013fqa,Haverkorn:2008tb}, the resulting conversion rate does not strongly depend on it~\cite{Ito:2023nkq}.
The solid and dashed red lines correspond to magnetic field magnitudes of $10\mu$G and $1\mu$G,
respectively.
As we will show in the next section, the graviton background produced in the Milky Way Galaxy is 
subdominant compared to that from blazar jets.

\section{Accumulated gravitational waves from blazar jets}\label{sec_blazar}
\begin{figure}[h]
\centering
\includegraphics[width=10cm]{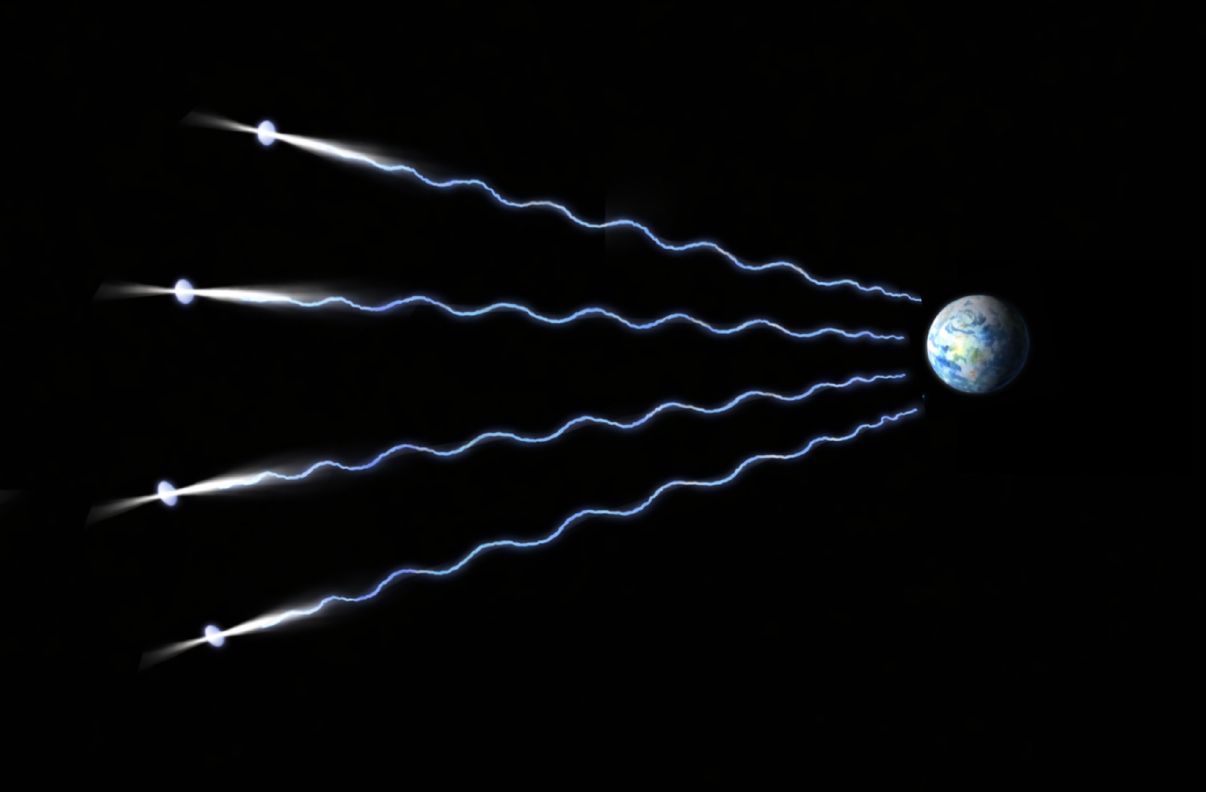}
\caption{Schematic illustration of the accumulated gravitational waves produced in individual blazar jets.
The disks represent the accretion disks, the white elongated structures represent the blazar jets,
and the wavy lines represent the gravitational waves produced through graviton-photon conversion
and propagating toward the Earth.}
\label{bla}
\end{figure}

In this section, we investigate the graviton-photon conversion within blazar jets 
and evaluate the generated total gravitational wave backgrounds around the Earth, 
as illustrated in Fig.\ref{bla}.

Blazar jets are relativistic outflows accompanied by magnetic fields, 
which play an important role in determining their emission properties.
It has been indicated that the helical component, which is almost orthogonal to jet direction, 
is dominant as is commonly observed in BL Lac objects~\cite{Marscher:2021ntl}.
This implies that graviton-photon conversion can occur efficiently during the propagation within the blazar jet.
In practice, simplified effective models of the magnetic field configuration are often adopted to fit the spectral energy distribution (SED) of the target blazar.
The one-zone synchrotron self-Compton (SSC) model and external
Compton (EC) model are often 
employed for this effective modeling~\cite{Tavecchio:1998xw,Paliya:2017xaq,Fan_2023}. 
In the SSC model, relativistic charged particles emit synchrotron radiation and subsequently upscatter these photons through inverse Compton scattering, all within a single emission region of radius $r_{{\rm em}}$, which is filled with a tangled and homogeneous magnetic field of strength $B_{{\rm em}}$.
It is known that the SSC model can fit the SEDs of BL Lac objects well.
On the other hand, the SEDs of flat spectrum radio quasars (FSRQs) are well described by 
the EC model, where the inverse-Compton component is produced by upscattering seed photons supplied from outside the emission region.
Outside the emission region, one would expect that magnetic fields expand freely without radiative pressure.
Thus, the strength of the magnetic field, envisioned as being dominated by its helical component, would scale as
\begin{align}
    B(r)&=B_{{\rm em}}  \hspace{2.25cm} \textrm{for}\enspace  r\leq r_{{\rm em}}, \label{B1}\\
    B(r)&= B_{{\rm em}}\left(\frac{r}{r_{{\rm em}}}\right)^{-1} \quad \textrm{for}\enspace r_{{\rm em}}<r . \label{B2}
\end{align}
Note that all quantities characterizing the jet models discussed here and in the following are measured in the comoving frame of the jet.
Since a jet exhibits relativistic bulk motion, with a Lorentz factor typically of $\Gamma\sim$ a few tens, 
the emission from the one-zone modeled jet is highly beamed with an opening angle of $\sim 1/\Gamma$ for an observer, although it is assumed to be isotropic in the jet’s comoving frame.
The $z$-axis used in Eqs.~(\ref{Ahplusmode}) and (\ref{Ahcrossmode}) corresponds to the line of sight and is taken to lie within the opening angle.
We assume that the above scaling of the magnetic field persists out to $r_{{\rm end}}$, 
where its amplitude becomes comparable to typical galactic magnetic fields, $\sim 1\mu{\rm G}$.
Below this field strength, the influence of the surrounding environment outside the jet may become significant, and the simple scaling may no longer be applicable.
%
%\begin{figure}[H]
%  \centering
%  \includegraphics[width=100.0truemm]{configuration.png}
%  \caption{three dimensional configuration}
%  \label{fig:3d-configuration}
%\end{figure}
%

In addition to the magnetic field configuration, the energy fraction carried by plasmas in the jet and their energy spectra must also be specified in order to reproduce the observed photon SED from the jet.
In this paper, we focus on the leptonic model, where
the emission is mainly attributed to synchrotron radiation and inverse Compton scattering by leptons only.
On the other hand, the hadronic model assumes that the high-energy component of SED, typically extending from X-rays to gamma rays, is dominated by hadron-initiated emission processes, such as proton synchrotron radiation and photomeson-induced cascades~\cite{Cerruti:2020lfj}.
%Since the hadronic model generally requires stronger magnetic fields than the leptonic model to reproduce the observed %SEDs, a higher graviton–photon conversion probability is achieved within jets~\cite{Matsuo:2025blj}.
%In this sense, the following estimate of the stochastic gravitational wave background
%based on the leptonic model is conservative.
Since the hadronic model generally requires stronger magnetic fields than the leptonic model
to reproduce the observed SEDs, a higher graviton-photon conversion probability can be achieved
within jets, as demonstrated in~\cite{Matsuo:2025blj}.
Although it would be interesting to evaluate the accumulated gravitational wave background
using the hadronic model, hadronic fits are not currently available for the full sample of 2708 blazars.
We therefore use the leptonic fits of~\cite{Fan_2023} for the full sample,
and the resulting gravitational wave background should be regarded as a conservative estimate.

We follow the leptonic model in~\cite{Fan_2023}, where
the energy distributions of electrons are taken 
to be
\begin{align}
    N(\gamma) &= N_{0}\left(\frac{\gamma}{\gamma_p}\right)^{-3} 10^{-2b\ln\left(\gamma/\gamma_p \right)} .
                 \label{N1}
\end{align}
Here $N_{0}$ is a normalization constant, $\gamma_p$ is a Lorentz factor contributing most to the synchrotron 
peak, and $b$ represents a constant corresponding to the synchrotron bump.
The distribution spans from the minimum Lorentz factor$\gamma_{{\rm min}}$ to the maximum Lorentz factor $\gamma_{{\rm max}}$.
Using Eq.(\ref{N1}), the 
averaged Lorentz factor is given by
\begin{align}
    \bar{\gamma}_i = \frac{\int\gamma N(\gamma) d\gamma}{\int N(\gamma) d\gamma}.
\end{align}
We assume that the ratio between the energy density of electrons, 
$U_e$, to that of the magnetic field, $U_B$, is constant.
This allows us to relate the distribution of the number densities of electrons to the magnetic field configuration as
\begin{align}
  n_e(r)&=\frac{B(r)^2}{2\bar{\gamma}_e m_e}\frac{U_e}{U_B} , \label{ne}
\end{align}
%

%Model parameters in the leptonic SSC or EC models can be determined by 
%fitting the SED of the blazar under study.
%In~\cite{Fan_2023}, the SEDs of 2708 blazar jets are analyzed, and model parameters 
%for each jet are determined.
%Using the parameter values obtained in~\cite{Fan_2023}, we calculate the graviton–photon conversion probability for %each jet, as photons produced in the emission region propagate from the outer edge of the emission region, 
%$r_{{\rm em}}$, to the end of the magnetic field $r_{{\rm end}}$ by numerically solving 
%Eqs.(\ref{Ahplusmode}) and (\ref{Ahcrossmode}).
Model parameters in the leptonic SSC or EC models can be determined by
fitting the SED of the blazar under study.
In~\cite{Fan_2023}, the SEDs of 2708 blazar jets are analyzed, and
the magnetic-field strength, emission-region size, electron Lorentz factor,
curvature of the electron energy distribution, and electron energy density
are determined for each jet.
Using these parameter values, we calculate the graviton-photon conversion
probability for each jet by numerically solving
Eqs.~(\ref{Ahplusmode}) and (\ref{Ahcrossmode}) as photons produced in the
emission region propagate from the outer edge of the emission region,
$r_{{\rm em}}$, to the end of the magnetic field, $r_{{\rm end}}$.
We note that thermal emission from accretion disks can also be important 
in the optical region of the SEDs of FSRQs.
Since most optical photons emitted from the accretion disk and reaching the observer 
pass through the emission region, 
their conversion after leaving the emission region can be estimated in the same manner.
%Note that the one-zone SSC modeling allows us to reduce the problem to one dimension, 
%as a consequence of averaging the magnetic field distribution in the jet’s comoving frame. 

As discussed in~\cite{Ito:2023fcr,Ito:2023nkq}, the conversion rate highly depends on the configuration of 
the magnetic field at the boundary, $r_{{\rm end}}$, if the oscillation length is short for the frequency 
under study, and even can be zero easily.
Therefore, we introduce cutoff frequencies by assuming that both very low and very high frequencies with short oscillation lengths have zero conversion rate, and calculate only the frequencies whose oscillation length at the boundary exceeds $r_{{\rm end}}$.
%Then, combining the SEDs~\cite{NASAIPAC} and the conversion probabilities, we evaluated the total 
%gravitational wave abundance around the Earth, which is given by the sum of the contributions from 
%all 2708 blazar jets.
Then, combining the observed photon SEDs taken from the NASA/IPAC
Extragalactic Database~\cite{NASAIPAC} with the conversion probabilities,
we evaluate the gravitational wave contribution from each source.
The total gravitational wave abundance around the Earth is obtained by summing
the contributions from all 2708 blazar jets.
\begin{figure}[th]
\centering
\includegraphics[width=12cm]{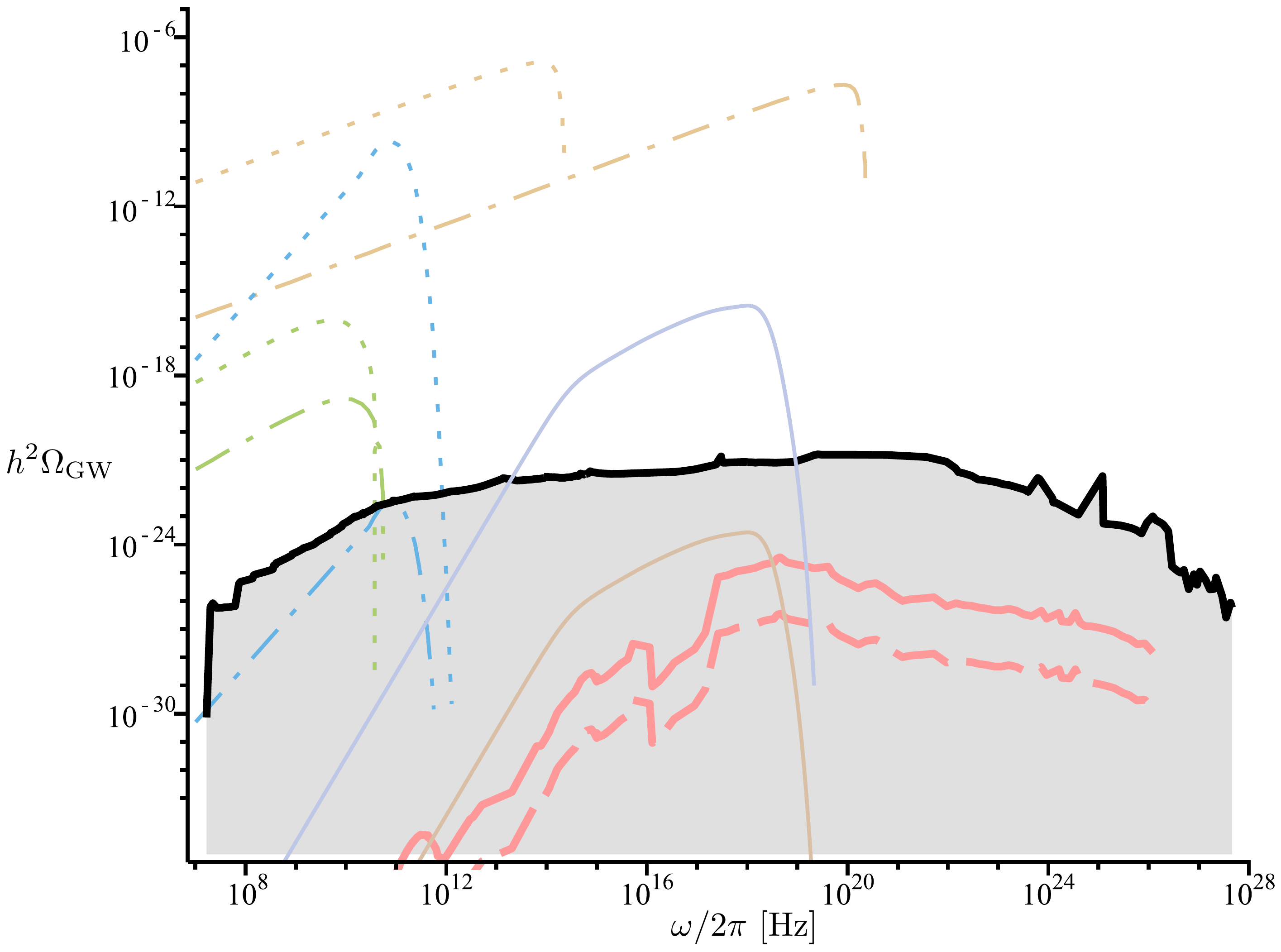}
\caption{The black line shows stochastic gravitational waves from graviton conversion in blazar jets.
The blue, orange, and green dotted (dot-dashed) lines correspond to gravitons produced from the thermal bath in the early universe with maximal temperatures of $10^{16}$GeV($10^{3}$GeV)~\cite{Aggarwal:2025noe,Ringwald:2020ist}, 
primordial black hole mergers with masses $10^{23}$g ($10^{17}$g)~\cite{Furuta:2025fbh}, 
and inflaton decay with quadratic (10th-order) potentials~\cite{Jiang:2024akb}, respectively.
The navy and brown lines represent gravitational waves from the Sun and 
main-sequence stars, respectively~\cite{Garcia-Cely:2024ujr}.
%The red line represents stochastic gravitational waves produced in the Milky Way Galaxy from the cosmic photon %background with the magnetic field magnitudes of $10\mu$G (see Fig\,.\ref{galaxy}).
The solid and dashed red lines represent stochastic gravitational waves produced in the Milky Way Galaxy
from the cosmic photon background with magnetic field magnitudes of $10\mu$G and $1\mu$G,
respectively (see Fig.~\ref{galaxy}).}
\label{blazar}
\end{figure}

The result is shown in Fig.~\ref{blazar} together with several other potential sources.
The black line represents stochastic gravitational waves from gravitons converted in each blazar jet.
The blue dotted (dot-dashed) line corresponds to graviton production from thermal bath in the early universe, extrapolated from the result in~\cite{Ringwald:2020ist},
where the maximal temperature is assumed to be $10^{16}$GeV ($10^{3}$GeV).
The orange  dotted (dot-dashed) lines represent 
stochastic gravitational waves from mergers of light primordial black holes (PBHs)~\cite{Furuta:2025fbh},
assuming that they constitute the entire dark matter abundance with a mass of 
$10^{23}$g ($10^{17}$g)~\cite{Carr:2009jm,Sasaki:2018dmp,Carr:2020gox}.~\footnote{See also Refs.~\cite{Wang:2016ana,Wang:2019kaf,Franciolini:2022htd,Kohri:2024qpd} for the other mass ranges of the merger signals. By considering the memory burden effect~\cite{Dvali:2020wft,Thoss:2024hsr,Chaudhuri:2025asm}, induced gravitational waves can be attractive sources for the high-frequency gravitational waves to test the PBHs to be dark matter~\cite{Kohri:2024qpd}.}
The green dotted (dot-dashed) lines represent stochastic gravitational waves from inflaton 
decay~\cite{Ema:2020ggo,Jiang:2024akb,Cline:2026jra}
in the post-inflationary universe with a quadratic (10th-order) polynomial inflaton potential~\cite{Jiang:2024akb}.
The navy and brown lines represent gravitational waves from the Sun and 
main-sequence stars, respectively~\cite{Garcia-Cely:2024ujr}.
Although the solar contribution may also constitute a graviton floor for high-frequency gravitational wave searches, it could be distinguished using directional information.
In contrast, the graviton floor obtained in this work arises from the accumulated contributions of 2708 blazars, and is therefore expected to be nearly isotropic.

%The red line represents stochastic gravitational waves produced in the Milky Way Galaxy from the cosmic photon %background, assuming 
%the magnetic field magnitudes of $10\mu$G and coherence length $100$pc, which is also 
%shown in Fig\,.\ref{galaxy}.
The red solid (dashed) line represents stochastic gravitational waves produced in the Milky Way Galaxy from the cosmic photon background, assuming
a magnetic field magnitude of $10\mu$G ($1\mu$G) and a coherence length of $100$pc, as also
shown in Fig.~\ref{galaxy}.
There are two main reasons why the graviton background produced in blazar jets dominates over that 
in the Milky Way Galaxy.
First, the magnetic field strength is larger in blazar jets.
Second, the electron energy can be ultra-relativistic, so that the plasma effect, 
which usually suppresses the conversion probability on the lower-frequency side, becomes smaller.
The shaded region below the black solid line indicates the region below the graviton floor.
For example, the blue dot-dashed line lies below the graviton floor, 
making the signal difficult to observe.

\section{Conclusion}
It has been observed that the Universe is permeated by a cosmic photon background spanning from radio waves to gamma rays~\cite{Hill:2018trh}, while magnetic fields are ubiquitous across the Universe, from small scales such as stars to galactic and even cosmological scales~\cite{Widrow:2002ud}.
We investigated the production of a graviton background generated through photon to graviton conversion 
in such magnetic fields.
In particular, we evaluated the graviton background produced in the Milky Way Galaxy in Sec.~\ref{Milky} 
and in blazar jets in Sec.~\ref{sec_blazar}.
It turned out that the dominant contribution to the graviton background arises from blazar jets, 
as shown in Fig.~\ref{blazar}.
Importantly, this graviton background constitutes a graviton floor for high-frequency gravitational wave detectors searching for new physics~\cite{Aggarwal:2025noe,Ito:2019wcb,Ito:2020wxi,Berlin:2021txa,Ito:2022rxn,Ito:2023bnu,Berlin:2023grv,Bringmann:2023gba,Kanno:2023whr,Ito:2025mgm,Cai:2025fpe}, 
analogous to the neutrino floor~\cite{Billard:2013qya,OHare:2021utq}.

We mention that the gravitational wave contribution from conversion in magnetic fields around objects such as the Earth, neutron stars, and white dwarfs is generally subdominant.
This is because such magnetic fields are localized and/or located far from us~\cite{Ito:2023fcr,Ito:2023nkq}.
As another possibility for generating the graviton background, magnetic fields in intergalactic regions may play a role.
However, there is a large uncertainty in the magnitude of the magnetic fields in current observations~\cite{MAGIC:2022piy,Jedamzik:2018itu,Planck:2015zrl,Paoletti:2022gsn,Pshirkov:2015tua},
so that the conversion probability also has a large uncertainty~\cite{Ito:2023nkq}.
Moreover, when the intergalactic magnetic fields were generated remains unknown.
This makes it difficult to establish a concrete graviton floor.
Therefore, it would be interesting to investigate this possibility once future observations better constrain the properties of intergalactic magnetic fields.

\section*{Acknowledgments}
We would like to thank Jaume Garriga for useful comments.
We would also like to thank Andreas Ringwald for helpful comments and for sharing data.
This work was supported by JSPS KAKENHI Grant Numbers JP26K17149 (AI), JP24K07027 (KK), and JP23K03411 (TS).

\bibliography{ref}% Produces the bibliography via BibTeX.

\end{document}